\documentclass[aps,pra,twocolumn]{revtex4}
\usepackage{graphicx}
\usepackage{dcolumn}
\usepackage{amsmath}
\usepackage{amssymb}
\usepackage{float}
\usepackage{epstopdf}

\def\exact{^{\rm exact}}

\def\pot{_{\rm pot}}

\def\ext{_{\rm ext}}

\def\ee{_{\rm ee}}
\def\xc{_{\rm xc}}
\def\pc{^{\rm PC}}
\def\isi{^{\rm ISI}}
\def\spl{^{\rm SPL}}
\def\glt{^{\rm GL2}}
\def\x{_{\rm x}}
\def\c{_{\rm c}}

\def\rv{{\bf r}}

\def\fv{{\bf f}}
\def\sv{{\bf s}}

\def\xv{{\bf x}}

\def\itPs{{\Psi}}
\def\beq{\begin{equation}}
\def\eeq{\end{equation}}

\def\nf{N_e}

\begin{document}
\title{Strictly correlated electrons in density functional theory:\\
A general formulation with applications to spherical densities}
\author{Michael Seidl}
\affiliation{Institute of Theoretical Physics,
University of Regensburg, D-93040 Regensburg, Germany}
\author{Paola Gori-Giorgi and Andreas Savin}
\affiliation{Laboratoire de Chimie Th\'eorique, CNRS,
Universit\'e Pierre et Marie Curie, 4 Place Jussieu,
F-75252 Paris, France}
\date{\today}

\begin{abstract}
We reformulate the strong-interaction limit of electronic density functional
theory in terms of a classical problem with a degenerate minimum. This
allows us to clarify many aspects of this limit, and to write
a general solution, which is explicitly calculated for spherical densities.
We then compare our results with previous approximate solutions and discuss
the implications for density functional theory.
 
\end{abstract}

\maketitle
\section{Introduction}
Density functional theory \cite{kohnnobel,science,FNM} (DFT)  is by now the
most popular method for electronic structure calculations in condensed matter
physics and quantum chemistry, because of its unique combination of low
computational cost and reasonable accuracy for many molecules and solids.

In applying DFT to a given electron system, the only quantity that
must be approximated in practice is the functional $E\xc[\rho]$ for the
exchange-correlation energy. An exact expression for this functional is
the coupling-constant integral,
\beq
E\xc[\rho]\;=\;\int_0^1 d\alpha\,W_{\alpha}[\rho].
\label{CCI}
\eeq
The integrand is defined as
\beq
W_{\alpha}[\rho]\;=\;\langle\Psi_{\alpha}[\rho]
        |\hat{V}\ee|\Psi_{\alpha}[\rho]\rangle -U[\rho].
\label{Walpha}
\eeq
Here, $U[\rho]=\frac12\int d\rv\int d\rv'\rho(\rv)\rho(\rv')/|\rv-\rv'|$
is the functional of the Hartree energy and the operator $\hat{V}\ee$
describes the Coulomb repulsion between the $N$ electrons,
\beq
\hat{V}\ee\;=\;\sum_{i=1}^{N-1}\sum_{j=i+1}^N\frac1{|\rv_i-\rv_j|}.
\label{Vee}
\eeq
(Atomic units are used throughout this work.) Eventually, out of all
antisymmetric $N$-electron wave functions $\Psi$ that are associated
with the same given electron density $\rho=\rho(\rv)$, $\Psi_{\alpha}[\rho]$
denotes the one that yields the minimum expectation of the operator
$\hat{T}+\alpha\hat{V}\ee$. Here, $\hat{T}=-\frac12\sum_{i=1}^{N}\nabla_i^2$
is the kinetic-energy operator. 
Notice that the parameter $\alpha$
works as an adjustable interaction strength or coupling ``constant''.

At $\alpha=0$, $W_{\alpha}[\rho]$ starts out with the value
\beq
E\x[\rho]\;=\;\langle\Psi_0[\rho]|\hat{V}\ee|\Psi_0[\rho]\rangle-U[\rho]
\label{Ex}
\eeq
which is the density functional for the exchange energy.
Generally, $W_{\alpha}[\rho]$ is a monotonically decreasing function
of $\alpha$, since the electrons in the state $\Psi_{\alpha}[\rho]$
increasingly tend to avoid each other in space as the repulsion
strength $\alpha$ grows. Thus, the expectation of $\hat{V}\ee$,
which is a measure for the average inverse distances between the electrons,
must decrease. However, it cannot decrease indefinitely, since the
electrons are forced to stay within the fixed density
$\rho(\rv)$. 

Schematically, the traditional quantum chemistry approach to electron
correlation often consists in trying to 
{\em extrapolate} the information on the physical system ($\alpha=1$)
from the non-interacting limit ($\alpha=0$) by using, e.g.,
perturbation theory or more sophisticated methods.
This work, instead, follows the early idea of Wigner \cite{wigner},
further developed in the DFT framework in Refs.~\onlinecite{SPL} and
\onlinecite{ISI}, in which the information 
at $\alpha=1$ is obtained
by {\em interpolating} between the two limits 
of weak interaction, $\alpha\to0$, and infinitely strong interaction, 
$\alpha\to\infty$.
In this context, we carry out a detailed study of the limit
\beq
W_\infty[\rho]\;=\;\lim_{\alpha\to\infty}W_\alpha[\rho].
\label{Winf}
\eeq
Although the limit of Eq. \eqref{Winf} was investigated in previous 
work \cite{SPL,SCE,PC,kieron},
the solution presented here was found only in the
special case of two electrons in a spherical density, using physical arguments.
For the general case with $N$ electrons, the point-charge-plus-continuum (PC) 
model was proposed \cite{SPL,PC},
\beq
W_\infty\pc[\rho]\;=\;\int d\rv\;\left[A\,\rho({\bf r})^{4/3}
   \;+\;B\,\frac{|\nabla\rho({\bf r})|^2}{\rho({\bf r})^{4/3}}\right].
\label{WPC}
\eeq
where $A=-\frac9{10}(\frac{4\pi}3)^{1/3}$     
and $B=\frac3{350}(\frac3{4\pi})^{1/3}$.      
This approximation, together with a similar
one for the next leading term, $W'_\infty[\rho]$, describing zero-point motion 
oscillations, was used to construct an interpolation for $W_\alpha[\rho]$
between $\alpha=0$ and $\alpha=\infty$, called interaction strength
interpolation (ISI) model \cite{ISI}. ISI predicts accurate atomization
energies (with a mean absolute error of 3.4 kcal/mole) \cite{ISI,ISI06},
showing that the general idea of interpolating between the weak- and the
strong-interaction limits in DFT can work.
 
It should be emphasized that ISI uses as ingredients exclusively the two
functionals $E\x[\rho]$ and $E\c\glt[\rho]$
(of the second-order correlation energy in
G\"orling-Levy perturbation theory \cite{GL}) from the relatively simple
non-interacting limit $\alpha\to0$
(with single-particle orbitals), plus the two functionals $W_\infty[\rho]$
and $W'_\infty[\rho]$ from the opposite $\alpha\to\infty$ limit of infinitely
strong repulsion, which are, so far, not known exactly
(as said, except for $W_\infty[\rho]$ in 
the special case of two electrons in a spherical density \cite{SCE}). 

In this work we reformulate the $\alpha\to\infty$ limit of DFT
in terms of a classical problem with a degenerate minimum. This allows us 
to construct the general solution for $W_\infty[\rho]$, and
to clarify many aspects of this limit.
As $\alpha>0$ grows beyond its real-world value $\alpha=1$,
the concept of single-particle orbitals becomes completely meaningless.
When $\alpha\to\infty$, however, an entirely new type of simplicity with
so-called ``co-motion'' functions arises, as we shall see in 
Secs.~\ref{sec_mr}-\ref{sec_fs}. The ``co-motion'' functions,
which entirely determine $W_\infty[\rho]$, can be
directly constructed from the one-electron density $\rho(\rv)$.

The paper is organized as follows.
In Sec.~\ref{sec_mr} we first give a general overview of the problem, 
anticipating the solution on intuitive physical grounds, and leaving
the mathematical details in
Secs.~\ref{subsec_as} and \ref{sec_fs}.
We then use our formalism to explicitly evaluate
the limit of Eq.~\eqref{Winf} in the case of spherical densities, 
with applications to  atoms (Sec.~\ref{sec_Nspherical}).
In Sec.~\ref{sec_isi} we compare our solution with the approximation
of Eq.~\eqref{WPC}, and we discuss the implications for the ISI functional.
The last Sec.~\ref{sec_conc} is devoted to conclusions and perspectives. In the Appendix
we also consider the simple case of harmonic forces, in order to analyze how
the nature of the electron-electron interaction affects the solution.

\section{Smooth densities from a classical problem}
\label{sec_mr}
For a given $N$-electron density $\rho=\rho(\rv)$ we generally wish
to find the limit \eqref{Winf} or, equivalently,
\beq
W_\infty[\rho]+U[\rho]=\lim_{\alpha\to\infty}
    \langle\Psi_{\alpha}[\rho]|\hat{V}\ee|\Psi_{\alpha}[\rho]\rangle.
\label{WinfDEF}
\eeq
If the density $\rho$ is both $N$- and $v$-representable for 
every $\alpha$, there
exists an $\alpha$-dependent external potential
$\hat{V}\ext^\alpha[\rho]=\sum_{i=1}^Nv\ext^\alpha([\rho],\rv_i)$
such that $\Psi_{\alpha}[\rho]$ is the ground state of the Hamiltonian
\beq
\hat{H}^\alpha[\rho]=\hat{T}+\alpha\hat{V}\ee+\hat{V}\ext^\alpha[\rho].
\label{Hv}
\eeq
As $\alpha\to\infty$, the binding external potential
$v\ext^\alpha([\rho],\rv)$ in the Hamiltonian \eqref{Hv} must compensate
the strong repulsion $O(\alpha)$ between the electrons. Therefore,
we expect
\beq
\lim_{\alpha\to\infty}\frac{v\ext^{\alpha}([\rho],\rv)}{\alpha}=v([\rho],\rv),
\label{eq_valphainf}
\eeq
with $v([\rho],\rv)$ a smooth function of $\rv$.
Thus,
for large $\alpha\gg1$, the kinetic energy in the state $\Psi_{\alpha}[\rho]$
is mainly due to zero-point oscillations of strongly repulsive electrons,
bound by a strong attractive force that has the order of $O(\alpha)$.
Therefore, $\langle\Psi_{\alpha}[\rho]|\hat{T}|\Psi_{\alpha}[\rho]\rangle$
has the order of $O(\sqrt{\alpha})$ and we may write in Eq.~\eqref{WinfDEF}
\begin{eqnarray}
W_\infty[\rho]+U[\rho]&=&\lim_{\alpha\to\infty}
\Big\langle\Psi_{\alpha}[\rho]\Big|\frac1{\alpha}\hat{T}+\hat{V}\ee\Big|
           \Psi_{\alpha}[\rho]\Big\rangle
\nonumber\\
&=&\lim_{\alpha\to\infty}\min_{\Psi\to\rho}
\Big\langle\Psi\Big|\frac1{\alpha}\hat{T}+\hat{V}\ee\Big|\Psi\Big\rangle.
\label{limmin}
\end{eqnarray}
In the second step we have applied the definition of the wave function
$\Psi_{\alpha}[\rho]$;
the constraint ``$\Psi\to\rho$'' addresses all those wave functions $\Psi$
that are associated with the same given density $\rho$.
Provided that the limit $\alpha\to\infty$ can be applied directly to the
operators in Eq.~\eqref{limmin} (there is no rigorous proof for this
reasonable conjecture -- see also the Appendix), it is simplified to
\beq
W_\infty[\rho]+U[\rho]\;=\;\min_{\itPs\to\rho}
\langle\itPs|\hat{V}\ee|\itPs\rangle.
\label{minc}
\eeq
In this case, the expectation of $\hat{V}\ee$ alone is to be minimized,
regardless of the kinetic-energy operator $\hat{T}$.
This apparently purely classical problem corresponds
to the quantum-mechanical limit of infinitely large masses. 

If Eq.~(\ref{eq_valphainf}) holds, that is if
the density $\rho(\rv)$ is both $N$- and $v$-representable also in the
very
$\alpha\to\infty$ limit, the minimizing $\Psi$ in Eq.~\eqref{minc} 
is the ground state of the pure multiplicative operator
$\hat{V}\ee+\hat{V}$.
The local one-body potential $\hat{V}$ is the Lagrange multiplier for the
constraint ``$\Psi\to\rho$'', and can be found via the 
Legendre-transform formulation of Eq.~\eqref{minc} \cite{maxmin},
\beq
W_\infty[\rho]+U[\rho]=\max_v\left\{\min_\Psi\langle\Psi|\hat{V}\ee+\hat{V}|
\Psi\rangle-\int \rho\,v\right\}.
\label{maxmin}
\eeq

To start to address the problem of $N$-representability in the
$\alpha\to\infty$ limit, 
we write the minimizing wavefunction 
$\Psi$ in Eq.~\eqref{maxmin} as the product
\beq
\Psi=\psi(\rv_1,...,\rv_N)\chi(\sigma_1,...,\sigma_N),
\label{eq_Psi}
\eeq
 where $\psi$ is a spatial wavefunction and $\chi$ is a chosen eigenstate 
of the total spin $\hat{S}^2$ and 
its projection $\hat{S}_z$. In Sec.~\ref{subsec_as} we analyze
the $N$-representability problem in more detail, 
showing that, in the special $\alpha\to\infty$ case, 
we can always construct an antisymmetric $\Psi$ of the
form \eqref{eq_Psi}, and that
the choice of the spin eigenfunction
$\chi$ does not affect the square of the spatial wavefunction 
$|\psi(\rv_1,...,\rv_2)|^2$ 
and thus, as shown by Eq.~\eqref{eq_psi2energy} below, the energy. 
For this reason, in what follows we only consider the
spatial wavefunction $\psi(\rv_1,...,\rv_N)$.

As said,
\beq
\hat{V}\ee+\hat{V}\equiv E\pot([v];\rv_1,...,\rv_N)
\label{opEpot}
\eeq
is a pure multiplicative operator, so that, after integrating out
the spin variables,
the unconstrained minimization in the brackets of Eq.~\eqref{maxmin} reads
\beq
\min_{\psi} \int d\rv_1...d\rv_N|\psi(\rv_1,...,\rv_N)|^2  
E\pot([v];\rv_1,...,\rv_N).
\label{eq_psi2energy}
\eeq
We see that the minimum in
Eq.~\eqref{eq_psi2energy} is reached
when the square of the spatial wavefunction $|\psi(\rv_1,...,\rv_N)|^2$ is a 
distribution that is zero everywhere except for values
$(\rv_1,...,\rv_N)\in M$, where
\beq
M\equiv\Big\{(\rv_1,...,\rv_N)\;:\;E\pot([v];\rv_1,...,\rv_N)=\mbox{min}\Big\}
\label{defM}
\eeq
is the set of all configurations $(\rv_1,...,\rv_N)$ for which the 
3$N$-dimensional function
$E\pot$ of Eq.~\eqref{opEpot} assumes its absolute (global) minimum. 
Notice that the set $M$ is purely
determined by the choice of $v(\rv)$, $M=M[v]$. We shall see in the
next Sec.~\ref{subsec_as} that to such a minimizing distribution
$|\psi(\rv_1,...,\rv_N)|^2$ we
can indeed associate an antisymmetric wavefunction of the form 
\eqref{eq_Psi}.

For a given reasonable (attractive) 
potential $v(\rv)$, we may expect that $M$ will 
comprise only one
single configuration (plus its permutations) or a small finite number
of configurations: the set $M$ corresponds, in fact, to the solution
of the classical electrostatic equilibrium problem for $N$ identical
point charges in the external potential $v(\rv)$.
The density associated with the corresponding spatial wave
function $\psi$ will then be a finite sum of delta functions,
\beq
\rho(\rv)\propto \sum_{k=1}^K \delta(\rv-\rv_k), 
\label{rho_delta}
\eeq
where the $\rv_k$ are the $K\ge N$ different position vectors forming the
configurations in $M$. When $v(\rv)$ has some symmetry (e.g., spherical),
$M$ can also be a low-dimensional continuum (e.g., the surface of a sphere).

For a quantum-mechanical system (atom, molecule, solid),
the electronic density is typically {\em smooth}, so that 
Eq.~(\ref{rho_delta}) seems in contradiction with the final requirement
``$\Psi\to\rho$'' [or the maximization \eqref{maxmin}]. In other
words, it seems like there is a $v$-representability problem for
 smooth densities in the very $\alpha\to\infty$ limit.
The solution to this apparent problem is that
the potential $v(\rv)$ must be such that 
$M[v]$ is a continuum with at least three dimensions. That is,
the absolute minimum of the $3N$-dimensional function
 $E\pot$ must be degenerate over  
(at least) a 3-dimensional subspace of ${\sf R}^{3N}$, which we write as
\beq
M=\Big\{\left(\rv,\fv_2(\rv),...,\fv_N(\rv)\right)\; : \;\rv\in P\Big\}
\label{Msol},
\eeq
where $P\subseteq {\sf R}^3$ is the region of space where $\rho(\rv)\neq 0$.
As we shall see in Sec.~\ref{sec_fs}, the ansatz \eqref{Msol} can be consistent
with the requirement that $E\pot$ is minimum over $M$ only
when the functions $\fv_n(\rv)$ satisfy special properties.

From the physical
point of view, the corresponding spatial wavefunction $\psi$ (which
is zero everywhere except on $M$), describes a state in which the position
of one of the electrons can be freely  chosen in $P$, $\rv_1=\rv$, but it then
 fixes the positions of all the other electrons through
the functions $\fv_n(\rv)$, $\rv_2=\fv_2(\rv),...,\rv_N=\fv_N(\rv)$.
This is what we call~\cite{SPL,SCE} ``strictly correlated electrons'' (SCE).

The requirement ``$\Psi\to\rho$'' can of course be fulfilled also 
if the degeneracy
of the absolute minimum of $E\pot$ is higher, e.g., if $M$ is a 6D subspace
described by $\{(\rv,\rv',\fv_3(\rv,\rv'),...,\fv_N(\rv,\rv'))\; : \;\{\rv,\rv'\}\in P\}$, which corresponds to the physical state in which the
position of two electrons fixes the positions of all the others. However,
we have to keep in mind that we can affect $M$ only by varying the 
one-body potential $v(\rv)$. Physically, this
means that $v(\rv)$ must compensate the repulsive forces of the other
$N-1$ electrons acting on the electron in $\rv$. 
If the position of the other $N-1$
electrons only depends on $\rv$ [as in Eq.~\eqref{Msol}], then it may
be possible
to find such a $v(\rv)$. If, instead, we require the degeneracy of $M$ to be
higher, finding such a $v(\rv)$ seems a daunting task. 

We shall go through these physical considerations
more in detail in Sec.~\ref{sec_fs}, where we analyze the 
solution~\eqref{Msol} and write down an explicit equation for $v(\rv)$.
Before doing so, we first show that we can consistently construct
an antisymmetric wavefunction $\Psi$ of the form \eqref{eq_Psi} such that
$|\psi(\rv_1,...,\rv_2)|^2$ is a distribution that is zero everywhere
except on the minimizing set $M$.

\section{Antisymmetry of the wavefunction} 
\label{subsec_as}
We want to construct an antisymmetric wavefunction of the form \eqref{eq_Psi}
such that its spatial part $|\psi(\rv_1,...,\rv_N)|^2$ is 
a distribution that is zero everywhere
except in the set $M$ where $E\pot$ has its global minimum. 
Given a configuration $(\rv_1,...,\rv_N)$ of $M$, we first notice 
that all its electronic positions $\rv_i$ 
must be different from each other: without
the kinetic energy operator $\hat{T}$, two or more electrons on-top of each 
other would make $E\pot$ infinite because of
the singular nature of Coulomb repulsion at short distances.
Moreover, for any configuration $\in M$, also all its $N!$ 
permutations are $\in M$,
since $E\pot$ is invariant under permutation of its variables $\rv_i$.
Then, given the value of the wavefunction for one configuration
$(\rv_1,...,\rv_N)\in M$, with spins $(\sigma_1,...,\sigma_N)$, its value
for the permutations $\pi$ of this configuration must be simply defined as
\begin{eqnarray}
\psi(\rv_{\pi(1)},...,\rv_{\pi(N)})\chi(\sigma_{\pi(1)},...,\sigma_{\pi(N)})
= \nonumber \\
(-1)^{\pi}\psi(\rv_1,...,\rv_N)\chi(\sigma_1,...,\sigma_N)
\label{eq_psianti}
\end{eqnarray}
where $(-1)^{\pi}$ denotes the sign of the permutation $\pi$. 
Because the $\rv_i$ in a given configuration are all different
from each other, and $\psi$ is zero everywhere except on $M$,
the $N$ electrons are always localized in different
regions of space that have exactly  zero overlap. This means that when
we consider the square of the spatial wavefunction 
 $|\psi(\rv_1,...,\rv_N)|^2$ all the cross terms are exactly zero,
and thus $|\psi|^2$ is
completely independent of the choice of the spin eigenstate $\chi$.

Similarly, for repulsive bosons (with any spin), 
$\Psi$ can be {\em symmetrized}. Both
the symmetric and the antisymmetric choice give the same spatial
$|\psi(\rv_1,...,\rv_N)|^2$, and thus the same expectation
for $\hat{V}\ee$: when Coulomb repulsion between the particles becomes
dominant with respect to the kinetic energy, particle-particle overlap is
suppressed, so that the particles no longer know wether they are fermions
or bosons.

The same construction \eqref{eq_psianti} applies to the SCE state in which
$M$ has the form \eqref{Msol}, since all the considerations made so far
are obviously still valid for any configuration in $M$ characterized
by a given $\rv\in P$.

\section{Co-motion functions}
\label{sec_fs}
We have seen that the potential $v(\rv)$ of
Eq.~\eqref{eq_valphainf} for the $\alpha\to\infty$
limit of DFT must be such that the minimum of the $3N$-dimensional
function $E\pot$ of Eq.~\eqref{opEpot} is degenerate over the 
3-dimensional subspace $M$ of Eq. \eqref{Msol}.  
We shall now analyze how
 to construct such a $v(\rv)$, and thus
which properties
the functions $\fv_n(\rv)$ must satisfy, and how they can be determined from
the density $\rho(\rv)$. Since the solution \eqref{Msol} corresponds
to a state in which the position of one electron dictates the postions
of all the others, we call the functions $\fv_n(\rv)$ {\em co-motion}
functions.

\subsection{Properties of the co-motion functions}
\label{propsfs}
If we assume that the $v(\rv)$ we are looking for is a smooth
potential with a continuous gradient $\nabla v(\rv)$, then
$E_{\rm pot}([v];\rv_1,...,\rv_N)$ is a continuous and smooth function
of its variables $\rv_n$, except for configurations with two or more electrons
on top of each other, which, as explained in Sec.~\ref{subsec_as},
 cannot belong to the minimizing set $M$.
In this case, a minimizing configuration
$(\rv_1,...,\rv_N)\in M$ must satisfy the stationarity conditions for
the function $E_{\rm pot}([v];\rv_1,...,\rv_N)$, i.e.,
\beq
\left\{
\begin{array}{l}
\nabla v(\rv_1)  =  \sum_{i\neq 1}^N \frac{\rv_1-\rv_i}{|\rv_1-\rv_i|^3} \\
\nabla v(\rv_2)  =  \sum_{i\neq 2}^N \frac{\rv_2-\rv_i}{|\rv_2-\rv_i|^3} \\
\vdots 
\end{array}
\right.
\label{gradients}
\eeq
If we insert the solution \eqref{Msol} into Eqs.~\eqref{gradients} we obtain
\beq
\left\{
\begin{array}{l}
\nabla v(\rv)  =  \sum_{i\neq 1}^N \frac{\rv-\fv_i(\rv)}{|\rv-\fv_i(\rv)|^3} \\
\nabla v(\xv)|_{\xv=\fv_2(\rv)}  =  \sum_{i\neq 2}^N \frac{\fv_2(\rv)-\fv_i(\rv)}{|\fv_2(\rv)-\fv_i(\rv)|^3} \\
\vdots 
\end{array}
\right.
\label{gradientswithf}
\eeq
where we have defined $\fv_1(\rv)\equiv \rv$. Now, 
suppose that we have found $N-1$
functions $\fv_i(\rv)$ and a potential $v(\rv)$ such that the first of the
Eqs.~\eqref{gradientswithf} is satisfied for all $\rv\in{\sf R}^3$.
If we now evaluate this first equation for $\rv=\fv_n(\sv)$ and then put
$\sv=\rv$, we see that its left-hand-side
coincides with the left-hand-side of the $n^{\rm th}$ equation. It is then
easy to verify that, if we want also the right-hand-sides to be the same,
the functions $\fv_i(\rv)$ must satisfy the transformation properties 
\beq
\{\fv_2(\fv_n(\rv)),...,\fv_N(\fv_n(\rv))\}=
\{\fv_1(\rv),...,\fv_N(\rv)\}\backslash \{\fv_n(\rv)\},
\label{propf}
\eeq
i.e., the set of $N-1$ functions $\fv_i(\xv)$ with $i=2,...,N$,
when evaluated in $\xv=\fv_n(\rv)$ must yield any permutation of the
set of $N-1$ $\fv_i(\rv)$ with $i=1,...,N$ and
$i\neq n$ (thus including $\fv_1(\rv)=\rv$).
This means that applying one co-motion function $\fv_k$ to one position $\rv_n$
of a given SCE configuration $C\in M$ must always yield a position of 
the same $C$ again. If the functions $\fv_i$ satisfy this property, then the
fulfillment of the first of the Eqs.~\eqref{gradientswithf} automatically
implies the fufillment of the other $N-1$ equations, and thus the
stationarity of the solution \eqref{Msol} for any $\rv$. One has then
to verify that such a stationary solution is the global minimum
of the $3N$-dimensional function $E\pot([v],\rv_1,...,\rv_N)$.

\subsection{The SCE external potential}
\label{sec_vSCE}
Once we have found some functions $\fv_i(\rv)$ that satisfy the
properties \eqref{propf},
Eqs.~\eqref{gradientswithf} provide $N$ equivalent equations for
the potential $v(\rv)$,
\beq
\nabla v(\rv)  =  \sum_{i\neq 1}^N \frac{\rv-\fv_i(\rv)}{|\rv-\fv_i(\rv)|^3}.
\label{eq_vSCE}
\eeq
Equation \eqref{eq_vSCE} has the clear physical meaning anticipated 
in Sec.~\ref{sec_mr}: the potential $v(\rv)$ must compensate the net
force acting on the electron in $\rv$, resulting from the repulsion
of the other $N-1$ electrons at positions $\fv_i(\rv)$.

We start now to see that the $\alpha\to\infty$ limit of DFT 
is entirely characterized by the co-motion functions $\fv_i(\rv)$,
which also give, as a byproduct via Eq.~\eqref{eq_vSCE}, the
external potential $v(\rv)$. In
the next Subsec.~\ref{sec_ffromrho} we shall see how to determine
these functions from the density $\rho(\rv)$.

\subsection{Co-motion functions for a given density}
\label{sec_ffromrho}
The SCE problem  for a given density $\rho(\rv)$ 
reduces then to the construction of the appropriate co-motion functions 
$\fv_n(\rv)$. To do this, we simply use the quantum mechanical
meaning of the electronic density. Since in the SCE state
the position of the first 
electron determines the positions
of all the others, the probability of finding the first electron in the volume element $d\rv$ around
the position $\rv$ must be the same of finding the $n^{\rm th}$ electron in the volume element
$d\fv_n(\rv)$ around the position $\fv_n(\rv)$. This means that all 
the co-motion functions $\fv_n(\rv)$ must satisfy the differential equation
\beq
\rho(\fv_n(\rv)) d\fv_n(\rv)=\rho(\rv) d\rv,\qquad n=2,...,N.
\label{detailedbalance}
\eeq
In order to construct the co-motion functions we thus have to find
the initial conditions for the
integration of \eqref{detailedbalance} that (i) satisfy the properties 
\eqref{propf}, (ii) yield a smooth potential $v(\rv)$ via
Eq.~\eqref{eq_vSCE}, and (iii) give the minimum
expectation of $\hat{V}\ee$. 
In Sec.~\ref{sec_Nspherical} we solve
this problem explicitly for spherically symmetric $N$-electron densities.

Equations \eqref{detailedbalance} can also
be proven by explicitly constructing the square of the spatial 
wavefunction $|\psi(\rv_1,...,\rv_N)|^2$
that is zero everywhere except on $M$ of Eq.~\eqref{Msol}, and by imposing
that the expectation value of the operator
$\hat{\rho}(\rv)=\sum_{i=1}^N\delta(\rv-\rv_i)$ on
$|\psi(\rv_1,...,\rv_N)|^2$ is $\rho(\rv)$. In this way,
we also see that the value of $|\psi(\rv,\fv_2(\rv)...,\fv_N(\rv))|^2$ in
a given SCE configuration $\in M$ must be set equal to
$\frac{1}{N}\rho(\rv)$.

\subsection{The value of $W_\infty[\rho]$}

After all, maximising with respect to the potential $v$ in
Eq.~\eqref{maxmin} seems to be equivalent to constructing the
co-motion functions $\fv_n(\rv)$.
In terms of these functions, the expectation of the operator
$\hat{V}\ee$ in the SCE state [see also Eqs.~\eqref{Walpha} and \eqref{Winf}] reads
\beq
W_\infty[\rho]+U[\rho]\;=\;\sum_{i=1}^{N-1}\sum_{j=i+1}^N
\int d\rv\,\frac{\frac1N\rho(\rv)}{|\fv_i(\rv)-\fv_j(\rv)|},
\label{WinfSCE}
\eeq
where, again, we have used the convention $\fv_1(\rv)=\rv$.
Equation~\eqref{WinfSCE} comes from the fact that, as said
in the previous Subsec.~\ref{sec_ffromrho},
for a given $\rv$, the 
configuration $(\rv,\fv_2(\rv),...,\fv_{N}(\rv))$ in $M$
has a weight equal to  $\frac1N\rho(\rv)$, and that 
the only possible electron-electron distances are
$|\fv_i(\rv)-\fv_j(\rv)|$, with $i\neq j$, and $i,j=1,...,N$.

\section{The SCE solution for $N$ electrons in a spherical density}
\label{sec_Nspherical}
As an example, we consider here $N$ electrons in a spherical density,
and we explictly calculate the functions $\fv_n(\rv)$ and 
the value $W_\infty[\rho]$ for a few atoms.

Given the symmetry of the problem, we choose a spherical-coordinate
reference system
$\rv_i=(r_i,\theta_i,\phi_i)$ in which the position of the
first electron defines the $z$-axis ($\theta_1=0,\phi_1=0$) and
the second electron is on the $xz$ plane ($\phi_2=0$). The $2N-3$
relative angles $\{\theta_2,\theta_3,\phi_3,...,\theta_N,\phi_N\}$ are
globally denoted by $\Omega$.
The total classical energy we want to minimize then reads
\beq
E_{\rm pot}([v],r_1,...,r_N,\Omega)=
\sum_{i=1}^N v(r_i)+V\ee(r_1,...,r_N,\Omega).
\eeq
Since the external potential does not depend on the relative angles $\Omega$,
we can decouple the stationarity equations \eqref{gradients} into an
angular part, which only involves $V\ee$, and a radial part
that also involves $v(r)$. By solving the angular part for any given
$r_1,...,r_N$ we can define a 
function $\Omega(r_1,...,r_N)$
for the minimizing angles, valid for any
spherically-symmetric $v$. These angles are the solution of the electrostatic
equilibrium problem for $N$ neutral sticks of lenghts $r_1,...,r_N$ having
the same point-charge $q$ glued at one end, and the other end fixed in the
origin, in such a way that they are free to rotate in the 3D space.

We now insert the function $\Omega(r_1,...,r_N)$
 in the stationarity equations for the radial variables $r_i$,
\beq
\left\{\begin{array}{l}
v'(r_1)=-\frac{\partial}{\partial r_1} V\ee(r_1,r_2,...,r_N,\Omega)
\big |_{\Omega(r_1,...,r_N)} \\
v'(r_2)=-\frac{\partial}{\partial r_2} V\ee(r_1,r_2,...,r_N,\Omega)
\big |_{\Omega(r_1,...,r_N)} \\
\vdots 
\end{array}
\right.
\label{grad_radial}
\eeq
As explained in the previous Sec.~\ref{sec_fs}, in order to have a smooth 
(e.g., atomic) radial density $\rho(r)$ we need to choose
a $v(r)$ that makes these equations fulfilled by a set
of radial distances $(r,f_2(r),...,f_N(r))$ in which $r$ can
take any value in the domain in which
$\rho(r)\neq 0$. These radial co-motion functions
$f_n(r)$ must have the same transformation properties
of Eq.~\eqref{propf}, as it can be verified by inspection of
Eqs.~\eqref{grad_radial}.

To construct the radial co-motion functions $f_n(r)$
for a given density $\rho(r)$ we proceed as follows. 
Our starting point is Eq.~\eqref{detailedbalance} that becomes, 
in the case of spherical symmetry,
\beq
4 \pi \,f_n(r)^2 \rho(f_n(r))\,|f'_n(r)|\,dr=4 \pi \,r^2 \rho(r)\,dr.
\label{dbspherical}
\eeq
We then define the function $\nf(r)$ that gives the expected number
of electrons between 0 and $r$,
\beq
\nf(r)=\int_0^r 4 \pi \,x^2 \rho(x)\,dx.
\eeq
$\nf(r)$ is a monotonously increasing function of $r$, and its inverse
$\nf^{-1}(y)$ is well defined for  $y\in[0,N)$. By integrating
both sides of Eq.~\eqref{dbspherical} we see that each function
$f_n(r)$ is of the form
\beq
f_n(r)=\left\{\begin{array}{ll}
\nf^{-1}(\nf(f_n(0))\pm \nf(r)) & 0\leq r\leq a_1 \\
\nf^{-1}(\nf(f_n(a_1))\mp \nf(a_1) \pm \nf(r)) & a_1\leq r\leq a_2 \\
\vdots
\end{array}
\right.
\label{fgeneral}
\eeq 
where the upper (lower) sign corresponds to the case $f'_n\geq 0$ ($f'_n<0$),
and $[0,a_1],\;[a_1,a_2],...,[a_{k-1},a_k]$ are the intervals in
which (i) $f'_n(r)$ does not change sign and (ii) $\nf(f_n(a_i))\mp 
\nf(a_i) \pm \nf(r)\in[0,N]$, so that $\nf^{-1}$ is well defined.

The SCE problem for spherical densities thus consists in finding the signs, 
the intervals
$[a_{i-1},a_i]$, and the initial conditions $f_n(0),...,f_n(a_k)$
in Eqs.~\eqref{fgeneral} that (i) make the $f_n(r)$ satisfy 
the transformation properties
\eqref{propf}, (ii) yield, via Eq.~\eqref{eq_vSCE},
a $v(r)$ with a continuous first derivative,
and (iii) give the minimum expectation
value of $\hat{V}\ee$ when inserted in Eq.~\eqref{WinfSCE}. 

We shall
now see how to proceed in the case of atomic densities. We first
treat the case of the He atom, and we then generalize the solution to
$N>2$.

\subsection{$N=2$ (He atom)}
\label{sub_He}
Two electrons in a smooth spherical density in the $\alpha\to\infty$
limit have been already considered in Ref.~\onlinecite{SCE}, where the SCE
state was proposed on the basis of physical considerations. Here, we
restate the problem in terms of the formalism just derived, and
we show that the solution proposed in Ref.~\onlinecite{SCE} was indeed
the correct one.

When $N=2$ we have only one relative angle, $\theta_2$. By solving
the stationarity equations for the angular part we immediatly obtain
$\theta_2=0$ or $\pi$. It is easy to verify that $\theta_2=0$ is a maximum
and $\theta_2=\pi$ is a minimum for any value $r_1$ and $r_2$. By
inserting the minimizing angle $\theta_2=\pi$ in the
stationarity equations for the radial variables, we obtain
\beq
\left\{\begin{array}{l}
v'(r_1)=(r_1+r_2)^{-2} \\
v'(r_2)=(r_1+r_2)^{-2} 
\end{array}
\right.
\eeq
These equations admit a solution of the kind $(r,f(r))$ if and
only if $f(r)$ satisfies the property $f(f(r))=r$, in agreement
with Eq.~\eqref{propf}. This means
that, among all the possible $f(r)$ of the form \eqref{fgeneral},
we can only choose the ones for which $f^{-1}=f$. 
One can verify that the only possible choices are then
\begin{eqnarray}
& & f(r)=\nf^{-1}(\nf(r))=r 
\label{breathing} \\
& & f(r)=\left\{\begin{array}{lr}
\nf^{-1}(\nf(f(0))-\nf(r)) & r<f(0) \\
\nf^{-1}(2+\nf(f(0))-\nf(r)) & r>f(0) \\
\end{array}
\right. 
\label{decreasing}
\\
& & f(r)=\left\{\begin{array}{lr}
\nf^{-1}(1+\nf(r)) & \qquad r<\nf^{-1}(1) \\
\nf^{-1}(\nf(r)-1) & \qquad r>\nf^{-1}(1) \\
\end{array}
\right. 
\label{increasing}
\end{eqnarray}
We call the choice \eqref{breathing} the ``breathing'' solution:
the two electrons are always at the same distance from the center, opposite
to each other. It is a stationary solution valid for any density, 
since $f(r)=r$ satisfies
the differential equation \eqref{dbspherical} independently of 
the choice of $\rho(r)$.
The corresponding external potential is $v(r)=-\frac{1}{4 r}$.
The Hessian matrix shows that this stationary
solution must be ruled out, since it 
is a maximum for $E_{\rm pot}([v],r_1,r_2,\theta_2)$.

The choice \eqref{decreasing} corresponds to a discontinuous 
$v'(r)$, except in the case $\nf(f(0))=2$, i.e., $f(0)=\infty$.
We rule out a discontinuous $v'(r)$ because otherwise 
our whole construction is inconsitent: if $v'(r)$ is not continuous
it is not necesessary for a minimizing configuration of
$E_{\rm pot}$ to satisfy the stationarity equations \eqref{gradients}.
As a double check, we also verified, in the case of the He
atom, that  the choice $\nf(f(0))=2$ in \eqref{decreasing} 
is indeed the one that yields
the lowest expectation of $\hat{V}\ee$, also when compared to
the form \eqref{increasing} that, again, corresponds to a discontinous $v'(r)$.
The SCE solution for $N=2$ electrons in a smooth
spherical density is thus the one proposed in Ref.~\onlinecite{SCE}, i.e.,
\beq
f(r)=\nf^{-1}(2-\nf(r)).
\label{fSCE2}
\eeq
It is easy to verify that this stationary solution, 
with its corresponding potential,
\beq
v(r)=\int^r\frac{dx}{(1+f(x))^2},
\eeq
is the true minimum of $E_{\rm pot}([v],r_1,r_2,\theta_2)$.

Besides the mathematical arguments, the choice \eqref{fSCE2} is the
most ``physical'' one: it makes the two electrons always be
in two different spherical shells, each one containing, on average in
the quantum mechanical problem, one electron~\cite{SCE}. 
This is exactly what we expect
for two electrons that repel each other infinitely strong, but that have to
fulfill the constraint of yielding a given smooth,
atomic-like density $\rho(r)$.

\subsection{$N=3$ (Li atom)}
When $N=3$ we have 3 angles, $\theta_2,\theta_3,\phi_3$, but
we immediately obtain $\phi_3=0$: the three electrons must be on the same
plane, containing the nucleus, to achieve compensation of
 the forces (see Sec.~\ref{sec_vSCE}). We then find
numerically, for any given $r_1,r_2,r_3$, the minimizing angles
$\theta_2$ and $\theta_3$.

\begin{figure}
\includegraphics[width=8.4cm]{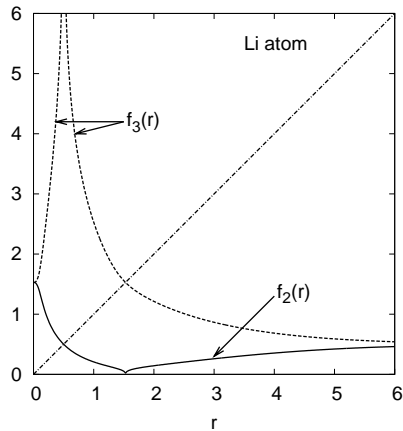} 
\caption{The radial co-motion functions
$f_2(r)$ and $f_3(r)$ for the Li atom density. Hartree atomic
units are used.}
\label{fig_f2f3Li}
\end{figure}

To construct the radial co-motion functions $f_2(r)$ and $f_3(r)$ 
we should, in principle, try out all the possible $f_n(r)$ of
the form \eqref{fgeneral} that satisfy the properties \eqref{propf},
and select the ones that yield a continous $v'(r)$ and the lowest
expectation for $\hat{V}\ee$. Instead, we use the physical idea
that was behind the solution for two electrons of Eq.~\eqref{fSCE2}:
we expect that the correct $f_2$ and $f_3$ are  the ones that make 
the three electrons
always occupy three different spherical shells, each containing, on average
in the quantum mechanical problem, one electron. Such radial co-motion
functions read
\begin{eqnarray}
& & f_2(r)=\left\{
\begin{array}{lr}
 \nf^{-1}(2-\nf(r)) & r\leq a_2 \\
  \nf^{-1}(\nf(r)-2) & r> a_2
\end{array}
\right. 
\label{f2Li} \\
& & f_3(r)=\left\{
\begin{array}{lr}
 \nf^{-1}(2+\nf(r)) & r\leq a_1 \\
  \nf^{-1}(4-\nf(r)) & r> a_1
\end{array}
\right.
\label{f3Li}
\end{eqnarray}
where $a_1=\nf^{-1}(1)$ and $a_2=\nf^{-1}(2)$. They are displayed,
for the Li atom using an accurate fully correlated density~\cite{bunge},
in Fig.~\ref{fig_f2f3Li}.

\begin{figure}
\includegraphics[width=8.0cm]{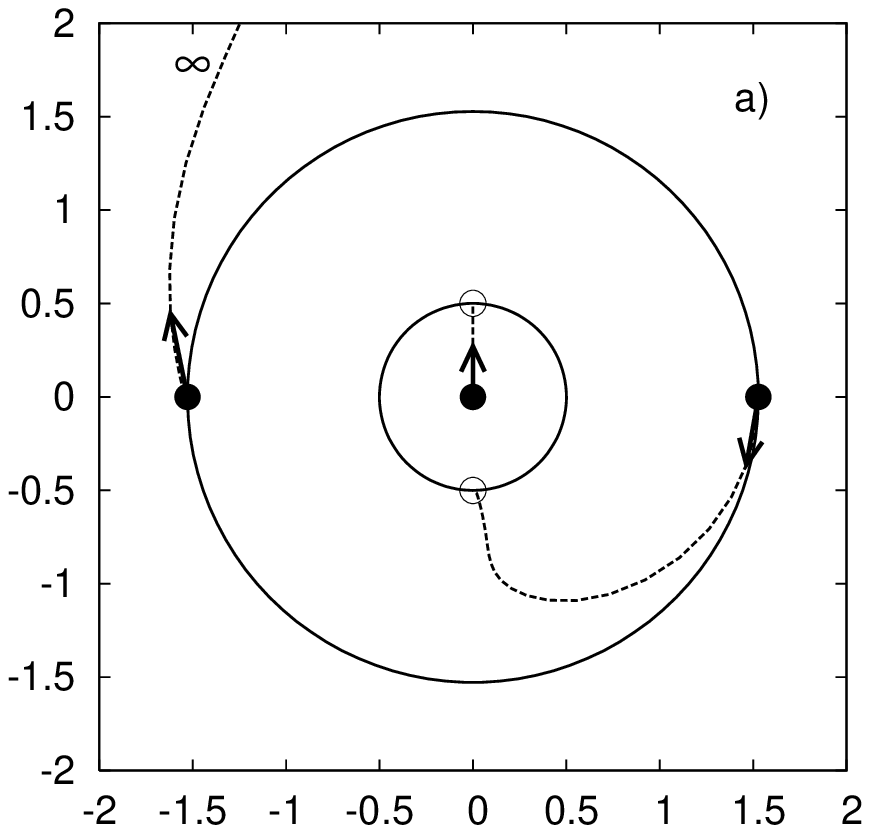} 
\includegraphics[width=8.0cm]{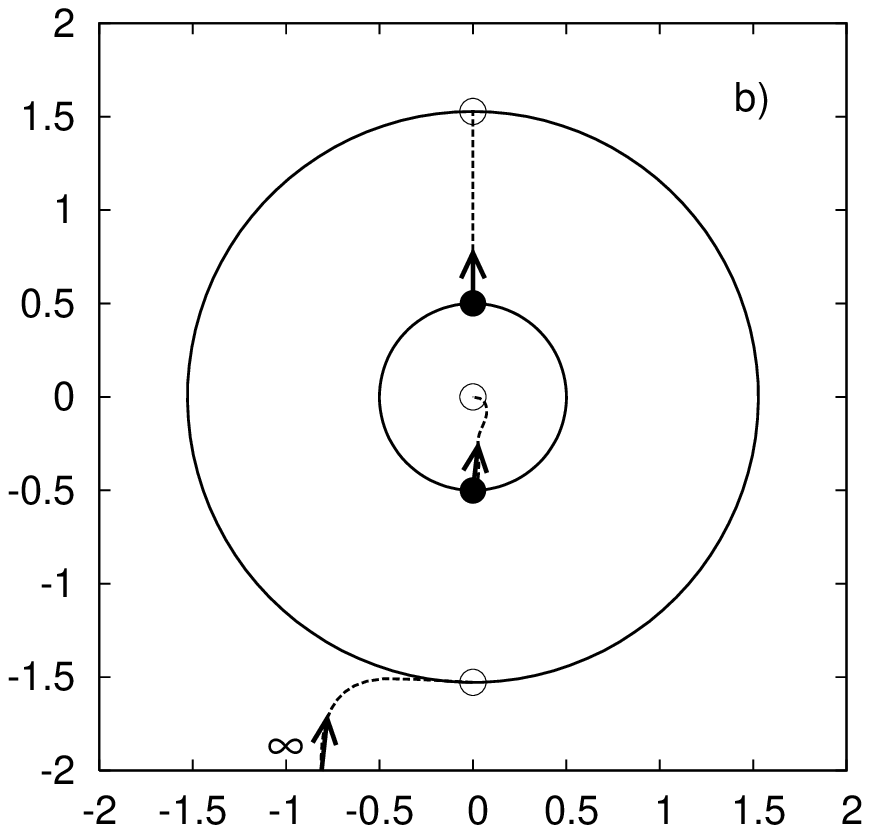} 
\includegraphics[width=8.0cm]{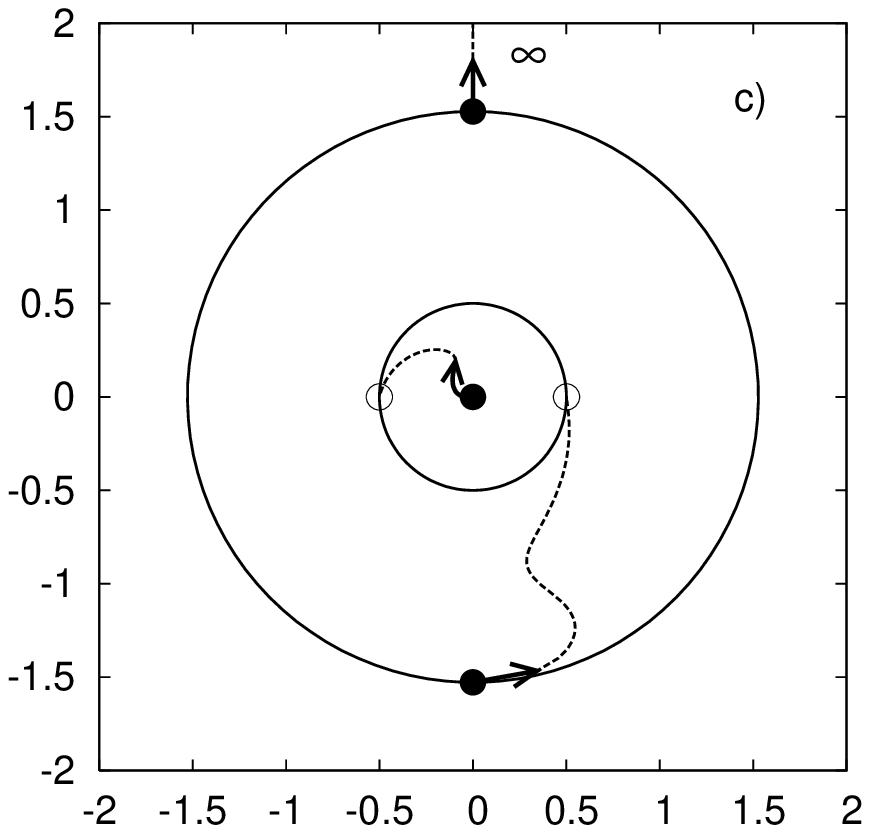} 
\caption{The positions of the 3 electrons in the SCE state for the
Li atom density. The two big circles
are the projections of the spheres containing, on average in the quantum
mechanical problem, 1 electron (radius $a_1$) and 2 electrons (radius $a_2$). The
position $r$ of the first electron is varied along the vertical $z$ axis
a) from 0 to $a_1$, b) from $a_1$ to $a_2$, and c) from $a_2$ to $\infty$.
The other two electrons have distances from the center given, respectively,
by $f_2(r)$ and $f_3(r)$ of Eqs.~\eqref{f2Li}-\eqref{f3Li}, 
and angular positions given by the 
minimizing angles $\theta_2(r,f_2(r),f_3(r))$ and  $\theta_3(r,f_2(r),f_3(r))$.
In each panel, the starting position of the 3 electrons is represented
by a full circle ($\bullet$) and their final position by an empty
circle ($\circ$). The dashed curves represent the ``trajectories'' 
of the 3 electrons, and the arrows their direction. The three panels
are actually equivalent (see text). Distances
are in atomic units.}
\label{fig_Lipositions}
\end{figure}

To better grasp the physics behind such solution, we show,
in Fig.~\ref{fig_Lipositions}, the positions of the three electrons
in the SCE state for the same accurate Li atom density~\cite{bunge}. 
The two big circles
are the radii of the spheres containing, on average in the quantum
mechanical problem, one electron ($a_1$) and two electrons ($a_2$). The
position $r$ of the first electron is varied along the vertical $z$ axis
a) from 0 to $a_1$, b) from $a_1$ to $a_2$, and c) from $a_2$ to $\infty$.
The other two electrons have distances from the center given, respectively,
by $f_2(r)$ and $f_3(r)$, and angular positions given by the
minimizing angles $\theta_2(r,f_2(r),f_3(r))$ and  $\theta_3(r,f_2(r),f_3(r))$.
In each panel, the starting position of the three electrons is represented
by a full circle ($\bullet$) and their final position by an empty
circle ($\circ$). The dashed curves represent the ``trajectories'' 
of the three electrons, and the arrows their direction. Thus, for each position
$r$  on the $z$ axis of the first electron, the positions of the other two 
electrons are
completely fixed, apart from the permutation symmetry of electron
2 with electron 3. We clearly see
from this figure, that in any SCE configuration the space is
divided in three spherical shells, which never contain more than one electron.
These three spherical shells are the same that, in the quantum mechanical
problem, contain {\em on average} one electron: the SCE state makes them
contain {\em always exactly } one electron, thus suppressing any accidental
clustering (``correlation suppresses fluctuations''~\cite{fulde,ziesche}). 

Because of the properties of the co-motion functions
(which are a consequence of the fact that the electrons are
indistinguishable), the three panels are equivalent.
Our physical problem is in fact invariant for a rigid rotation
of any of the configurations corresponding to a given $r$. This means
that in panel b) we could, for each value of $r$, rigidly rotate the
position of the three electrons in such a way that the electron in
the inner shell moves on a straight line. In this way, we would reobtain
a rigid rotation of panel a). The same can be done with panel c). Thus,
to calculate the expectation of $\hat{V}\ee$ we only need to consider
the case $0\leq r\leq a_1$.

The solution of Eqs.~\eqref{f2Li}-\eqref{f3Li} is thus very reasonable
from the physical point of view. Moreover, it can be easily verified 
that it satisfies
the properties required by stationarity of Eq.~\eqref{propf}, and that
it corresponds to a smooth external potential $v(r)$, which
can be computed via
\beq
v'(r)=-\frac{\partial}{\partial r_1} V\ee(r_1,r_2,r_3,\theta_2,\theta_3)
\big |_{r,f_2(r),f_3(r),\theta_2(r),\theta_3(r)},
\label{potfromvee3}
\eeq
where $\theta_i(r)$ is a shortened notation for $\theta_i(r,f_2(r),f_3(r))$.
\begin{figure}
\includegraphics[width=7.8cm]{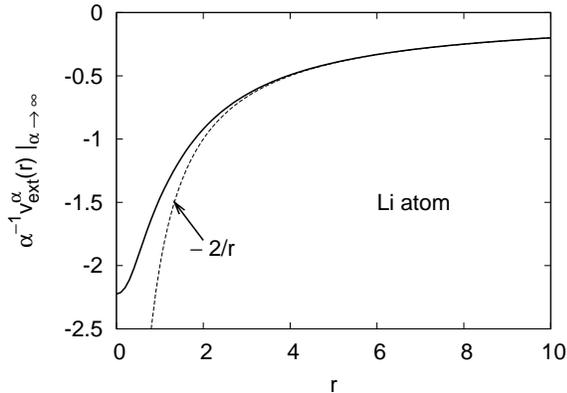} 
\caption{The external potential $v(r)$ of Eq.~\eqref{eq_valphainf}
for the SCE state in the case
of the Li atom. The asymptotic expansion for large
$r$, $v(r\to\infty)=-\frac{N-1}{r}$, is also
reported. All quantities are in Hartree atomic units.}
\label{fig_Lipot}
\end{figure}
The potential obtained by integrating Eq.~\eqref{potfromvee3} 
(again using
the accurate density for the Li atom of Ref.~\cite{bunge}) 
is reported in Fig.~\ref{fig_Lipot}. Notice that $v(r)$ is finite, with 
a zero first derivative, at $r=0$: the electron-nucleus cusp in the density
$\rho(r)$ is compensated by the kinetic energy for any value
of $\alpha$ (including the very limit $\alpha\to\infty$), 
as discussed in Ref.~\onlinecite{nagy}. For large $r$, we have
$v(r\to\infty)= -\frac{N-1}{r}$, a necessary condition to 
include as a stationary configuration the one in which one of the $N$
electrons is at infinity. The large-$r$ asympotic expansion $-2/r$ is also
shown in the same Fig.~\ref{fig_Lipot}.

With this potential $v(r)$ we have also verified
that the stationary solution of Eqs.~\eqref{f2Li}-\eqref{f3Li} is 
the absolute minimum of
$E_{\rm pot}([v],r_1,r_2,r_3,\theta_2,\theta_3,\phi_3)$. All these checks
make us believe that Eqs.~\eqref{f2Li}-\eqref{f3Li} are the correct
SCE solution for the Li atom: it is  difficult to imagine a different
solution that yields a lower value for the expectation of $\hat{V}\ee$,
since our solution is the one that makes the 3 electrons be always as
far as possible from each other, 
without violating the constraint of yielding the smooth density
$\rho(r)$. 

We can thus finally obtain the value 
$W_{\infty}[\rho]=-2.6030$~Hartree for the Li atom, via
\begin{eqnarray}
W_{\infty}[\rho] + U[\rho]=\nonumber \\
 \int_0^{a_1} dr\,4\pi\,r^2\rho(r)\,V\ee(r,f_2(r),
f_3(r),\theta_2(r),\theta_3(r)), 
\label{WinftyLi}
\end{eqnarray}
where we have used the fact that, because the three electrons are
indistinguishable,
integrating from 0 to $\infty$ is the same as integrating three times from
0 to $a_1$. 

As a final remark, we comment briefly on the fact that one never
obtains the configuration in which the three electrons are at the
same distance from the nucleus, at the vertices of an equilateral triangle.
The reason is that the only $f_2(r)$ and $f_3(r)$ compatible with
this configuration (even for only one single value of
$r$, say $r_0$) that also satisfy the required properties \eqref{propf}
are the ones corresponding to the ``breathing'' solution,
$f_2(r)=f_3(r)=r$, as it can be easily verified by integrating both
sides of Eq.~\eqref{dbspherical} with the initial condition $r=r_0,\;
f_n(r_0)=r_0$. This gives the two possibilites $f_A(r)=r$ or
$f_B(r)=\nf^{-1}(2\nf(r_0)-\nf(r))$: any choice for $f_2(r)$ and $f_3(r)$ 
which contains the function $f_B(r)$ does not satisfy the properties
\eqref{propf}. This is a difference with the $N=2$ case in which, instead,
the solution is $f(r)=f_B(r)$ with $\nf(r_0)=1$: two electrons in the
SCE state in a smooth density
have one configuration in which they are both at the same
distance from the nucleus, but for three electrons this does not happen.
One can also easiliy check that, again, the ``breathing'' solution, with
its potential $v(r)=-\frac{1}{\sqrt{3} r}$, is not a minimum
for the corresponding $E_{\rm pot}([v],r_1,r_2,r_3,\theta_2,\theta_3,\phi_3)$.

We can at this point clearly see the difference between the
SCE state for a smooth atomic density (Fig.~\ref{fig_Lipositions}) and
the more familiar Wigner-crystal-like state (for a case with
 three electrons see, e.g.,
Ref.~\onlinecite{kasia}), in which the three electrons are
localized at the vertices of a triangle at a certain distance from the nucleus,
say $r_0$. 
In
this latter case, the density becomes very peaked around $r_0$, 
loosing any resemblence with an atomic density, and becoming
more and more similar to Eq.~\eqref{rho_delta}.

\subsection{$N=4$ (Be atom)}
We have now  5 relative angles to consider. As in the $N=3$ case, 
we compute the minimizing
angular function $\Omega(r_1,r_2,r_3,r_4)$ numerically.

\begin{figure}
\includegraphics[width=8.4cm]{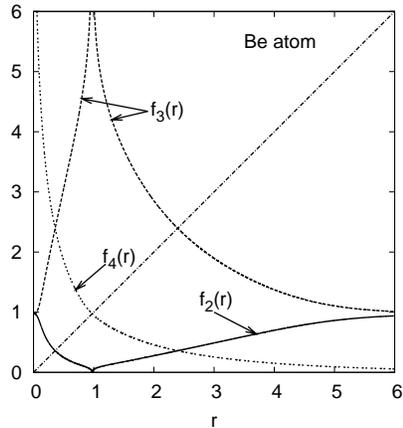} 
\caption{The radial co-motion functions
$f_2(r)$, $f_3(r)$ and $f_4(r)$ for the Be atom density. Hartree atomic
units are used.}
\label{fig_fBe}
\end{figure}

We then construct the radial co-motion functions, $f_2(r)$, $f_3(r)$,
$f_4(r)$, following the same ideas used for the case $N=2$ and 3: we divide
the space in four spherical shells, each containing, on average in the
quantum mechanical problem, one electron. The radial co-motion
functions that make the four electrons always be in four distinct
shells are then
\begin{eqnarray}
& & f_2(r)=\left\{
\begin{array}{lr}
 \nf^{-1}(2-\nf(r)) & r\leq a_2 \\
  \nf^{-1}(\nf(r)-2) & r> a_2
\end{array}
\right. 
\label{f2Be} \\
& & f_3(r)=\left\{
\begin{array}{lr}
 \nf^{-1}(2+\nf(r)) & r\leq a_2 \\
  \nf^{-1}(6-\nf(r)) & r> a_2
\end{array}
\right.
\label{f3Be} \\
& & f_4(r)=\nf^{-1}(4-\nf(r)),
\label{f4Be}
\end{eqnarray} 
where, again, $a_i=\nf^{-1}(i)$. These functions satisfy 
the transformation properties \eqref{propf}, and are reported in
Fig.~\ref{fig_fBe} for the case of an accurate correlated
density~\cite{umrigar} of the Be atom. 

We can then obtain, using the analogue of Eq.~\eqref{WinftyLi} for
$N=4$, the value $W_\infty[\rho]=-4.0212$ Hartree for the Be atom.

\begin{figure}
\includegraphics[width=7.8cm]{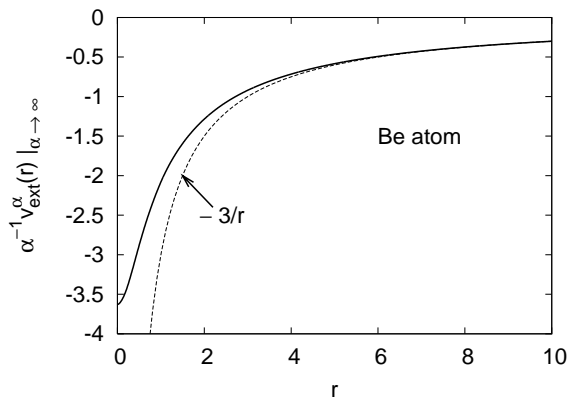} 
\caption{The external potential $v(r)$ of Eq.~\eqref{eq_valphainf}
for the SCE state in the case
of the Be atom. The asymptotic expansion for large
$r$, $v(r\to\infty)=-\frac{N-1}{r}$, is also
reported. All quantities are in Hartree atomic units.}
\label{fig_Bepot}
\end{figure}

To check our result, we also computed the corresponding external
potential $v(r)$, which, as shown in Fig.~\ref{fig_Bepot}, is continuous and
smooth. We also verified that our stationary solution is the absolute minimum
for the corresponding $E_{\rm pot}([v],r_1,r_2,r_3,r_4,\Omega)$.

Interestingly, we found that, for the case of the Be atomic density,
the four electrons always lie on the same plane, containing the nucleus. 
This effect seems to be
due to the shell structure of the atomic density: by repeating
the calculation using a simple exponential
density, $\rho(r)=\beta^3 e^{-\beta\,r}/2\pi$, we found that the
four electrons are often non-complanar. Instead, by using
other test densities with a shell structure, we found again the four electrons
always on the same plane. The shell structure seems to force the four
electrons to have distances from the nucleus that are distributed in a highly
nonuniform way, so that forces compensation (see Sec.~\ref{sec_vSCE}) 
can be achieved only if they are on the same plane.

\subsection{The general $N$-electron solution}
This way of constructing the solution can be generalized to any
number of electrons $N$. The radial co-motion functions that make
 the $N$ electrons
always occupy $N$ different shells containing, on average in
the quantum mechanical problem, one electron, are as follows.
Define an integer index $k$ running for odd $N$ from 1 to $(N-1)/2$, 
and for even $N$ from 1 to $(N-2)/2$. Then
\begin{eqnarray}
 f_{2k}(r)=\left\{
\begin{array}{lr}
 \nf^{-1}(2k-\nf(r)) & r\leq a_{2 k} \\
  \nf^{-1}(\nf(r)-2k) & r> a_{2k}
\end{array}
\right. 
 \\
 f_{2k+1}(r)=\left\{
\begin{array}{lr}
 \nf^{-1}(\nf(r)+2k) & r\leq a_{N-2 k} \\
  \nf^{-1}(2N-2k-\nf(r)) & r> a_{N-2k},
\end{array}
\right.
\end{eqnarray}
where  $a_i=\nf^{-1}(i)$.
For odd $N$, these equations give all the needed $N-1$ radial co-motion 
functions, while for even $N$ we have to add the last function,
\beq
f_N(r)= \nf^{-1}(N-\nf(r)).
\eeq

\begin{figure}
\includegraphics[width=8.4cm]{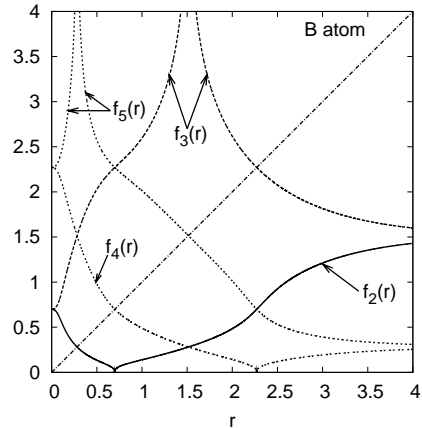} 
\caption{The radial co-motion functions
for the sphericalized B atom density. Hartree atomic
units are used.}
\label{fig_fB}
\end{figure}

\begin{figure}
\includegraphics[width=8.4cm]{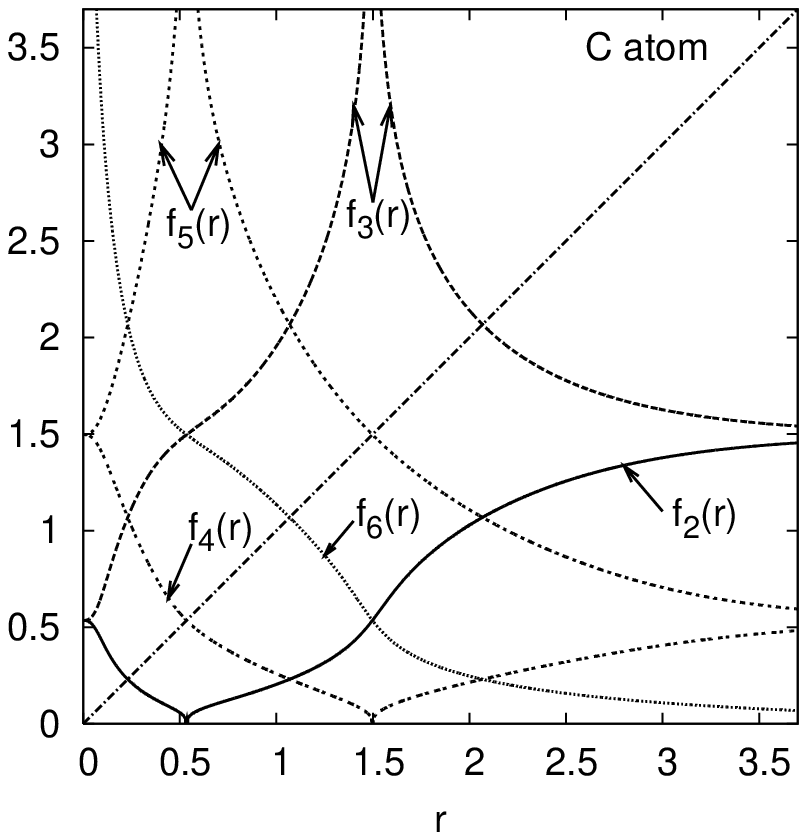} 
\caption{The radial co-motion functions
for the sphericalized C atom density. Hartree atomic
units are used.}
\label{fig_fC}
\end{figure}

\begin{figure}
\includegraphics[width=8.4cm]{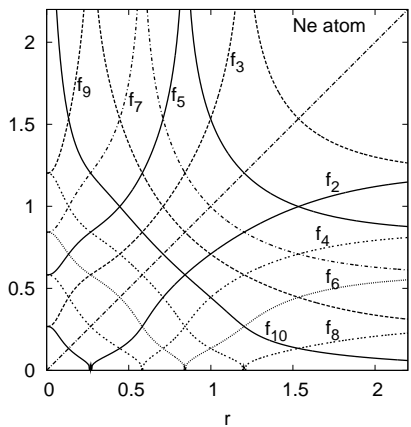} 
\caption{The radial co-motion functions
for the Ne atom density. Hartree atomic units are used.}
\label{fig_fNe}
\end{figure}

\begin{figure}
\includegraphics[width=7.8cm]{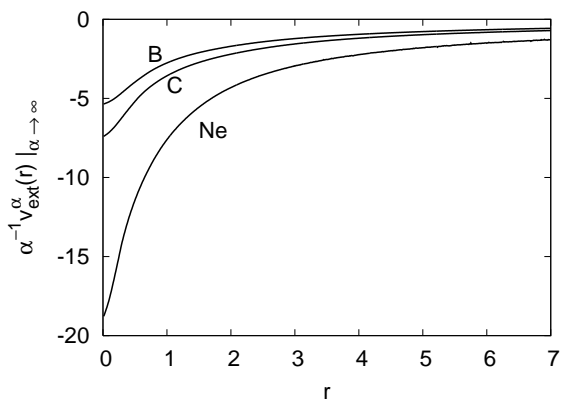} 
\caption{The external potential $v(r)$ of Eq.~\eqref{eq_valphainf}
for the SCE state in the case
of the Ne atom and of the sphericalized B and C atoms. 
All quantities are in Hartree atomic units.}
\label{fig_BCNepot}
\end{figure}

Using these radial co-motion functions, 
which satisfy the properties \eqref{propf} for any $N$, we calculated the 
SCE value $W_{\infty}[\rho]$ for accurate sphericalized densities of
the B and C atoms~\cite{julien}, and for the Ne 
atom density \cite{umrigar,umrigarNe}. The resulting radial
co-motion functions are displayed in Figs.~\ref{fig_fB}-\ref{fig_fNe}.
The results for $W_{\infty}[\rho]$  are reported
in Table~\ref{tabWPC}, and the corresponding potentials 
$v(r)$ are shown in Fig.~\ref{fig_BCNepot}. In all cases
we find the $N$ electrons non-complanar.

\section{Comparison with the PC model}
\label{sec_isi}
In the absence of spherical symmetry there seems to be no general
strategy for finding the initial conditions to integrate
Eqs. \eqref{detailedbalance}, although
we are still investigating this problem. So far, such cases
can only be treated approximately, using instead of the exact expression
\eqref{WinfSCE} the PC model \eqref{WPC} for the functional $W_\infty[\rho]$.
The PC model has been tested successfully \cite{PC} against the meta-GGA
functional of Ref.~\onlinecite{PKZB} that is expected to 
be accurate in the limit
$\alpha\to\infty$. The present exact (SCE) results, in contrast, 
provide a rigorous test for the PC model.

Table \ref{tabWPC}
documents reasonable values from the PC model 
for all the atoms considered here. A severe
test is the trivial case of the hydrogen atom H, with the one-electron density
$\rho(r)=\frac1{\pi}e^{-2r}$, where $\langle\hat{V}\ee\rangle=0$ for any value
of $\alpha$ and thus $W_\infty[\rho]=-\frac5{16}$ Hartree.
For the other atoms in Table \ref{tabWPC},
the PC error is less than 63 mH $\approx$ 39 kcal/mole. 

\begin{table}
\begin{tabular}{lllc}
\hline \hline
& $W_\infty^{\rm SCE}[\rho]$ (H) & $W_\infty^{\rm PC}[\rho]$ (H) & error (mH)\\
\hline 
H  & $-0.3125$ & $-0.3128$ & 0.3\\
He & $-1.500$  & $-1.463$ & 37 \\
Li & $-2.6030$ & $-2.5559$ & 47\\
Be & $-4.0212$ & $-3.9608$ & 60\\ 
B &  $-5.7063$ & $-5.6502$ & 56 \\
C &  $-7.7817$ & $-7.7192$ & 63\\ 
Ne & $-20.035$ & $-20.000 $ & 35\\ 
\hline \hline
\end{tabular}
\caption{Comparison of the values $W_\infty[\rho]$ in Hartree atomic
units obtained
with the SCE construction, and with the PC model \cite{PC}. 
The absolute errors of the PC model are also reported.
}
\label{tabWPC}
\end{table}
The quantity $W_\infty[\rho]$ refers to the limit of infinitely strong
repulsion between electrons. Being the subject of the present paper,
this extreme limit is not directly relevant for real electron systems
such as atoms or molecules. Rather, it is studied here in order to be combined
with additional information from the opposite weak-interaction limit
$\alpha\to0$. In this way, properties of real atoms and molecules, 
including those with weak correlations, can be predicted
accurately \cite{ISI} without dealing explicitly with the complicated
wave function at the realistic interaction strength $\alpha=1$. Both
the extreme limits $\alpha\to0$ and $\alpha\to\infty$ are mathematically
much simpler than the realistic situation at $\alpha=1$.
Consequently, 
more relevant than the error of the quantity $W_\infty[\rho]$ itself is the
error it causes in the ISI correlation energy \cite{ISI}
$E\c\isi[\rho]=E\xc\isi[\rho]-E\x[\rho]$.
Since the ISI model of Ref.~\onlinecite{ISI} uses, along with 
$W_\infty[\rho]$, also the
coefficient $W'_\infty[\rho]$ which is not known exactly, we consider here
the earlier ISI version SPL of Ref.~\onlinecite{SPL},
\beq
W_{\alpha}\spl[\rho]\;=\;
W_{\infty}[\rho]+\frac{E\x[\rho]-W_{\infty}[\rho]}{\sqrt{1+2Q[\rho]\,\alpha}},
\label{WSPL}
\eeq
with
\beq
Q[\rho]=\frac{2|E\c\glt[\rho]|}{E\x[\rho]-W_{\infty}[\rho]}.
\eeq
Analytical integration of $\int_0^1d\alpha W_{\alpha}\spl[\rho]-E\x[\rho]$ yields
\beq
E\c\spl[\rho]=\Big(E\x[\rho]-W_{\infty}[\rho]\Big)
\left[\frac{\sqrt{1+2Q[\rho]}-1}{Q[\rho]}-1\right].
\eeq
The correlation energies $E\c\spl[\rho]$, 
evaluated with the values $W_\infty[\rho]$
of Table \ref{tabWPC}, are reported in the columns ``(PC)'' and ``(SCE)''
of Table \ref{tabSPL}. 
While the SPL prediction comes
quite close to the exact correlation energy of the He atom, the case of
the Be and Ne atoms is less satisfying. Nevertheless, the improvement beyond
the second-order correlation energy  is remarkable,
since the only additional information used here is the coefficient
$W_\infty[\rho]$ (but not $W'_\infty[\rho]$).
\begin{table}
\begin{tabular}{llllll}
\hline\hline
& $E\x[\rho]$ & $E\c\glt[\rho]$ & (PC) & (SCE) & $E\c\exact[\rho]$ \\
\hline 
He & $-$1.0246 & $-$0.0503 & $-$0.0413 & $-$0.0418 & $-$0.042 \\
Be & $-$2.674  & $-$0.125  & $-$0.1054 & $-$0.1061 & $-$0.096 \\
Ne & $-$12.084 & $-$0.469  & $-$0.4205 & $-$0.4207 & $-$0.394 \\
\hline\hline
\end{tabular}
\caption{The exchange energy $E\x[\rho]$, the second order G\"orling-Levy
correlation energy $E\c\glt[\rho]$ \cite{ernz}, and the
estimate of the correlation energy $E\c[\rho]$ from the SPL
ISI model \cite{SPL} using the values $W_\infty^{\rm SCE}[\rho]$ (SCE) 
and $W_\infty^{\rm PC}[\rho]$ (PC), compared with ``exact'' values. 
All energies are in Hartree atomic units.}
\label{tabSPL}
\end{table}

For the He, Be, and Ne atoms, the error introduced by the PC model 
in $E\c\spl[\rho]$ is always
less than 1 mH, showing that in the case of neutral atoms
the SPL-ISI correlation functional is not too sensitive to the exact value of
$W_{\infty}[\rho]$. However, we have to keep in mind that neutral atoms
are systems that resemble much more to the noninteracting KS system 
($\alpha=0$) than to the SCE state ($\alpha=\infty$). We expect that
for more correlated systems (e.g., streched bonds) the ISI correlation
energy is much more sensitive to the exact value of $W_{\infty}[\rho]$.
The investigation of such cases will be the object of future work.

\section{Conclusions and perspectives}
\label{sec_conc}
We have reformulated the strong-interaction limit of density functional
theory in terms of a classical problem with a degenerate minimum, obtaining
 a consistent solution for this
limit. Even if an antisymmetric wavefunction can be explicitly 
constructed, the strong-interaction limit of DFT 
is entirely characterized by much simpler
quantities, the so called ``co-motion'' functions, which are related
to the electronic density $\rho(\rv)$ via the differential 
equation \eqref{detailedbalance}. The results of this work can be useful
to construct interpolations between the weak- and the strong-interaction
limits of DFT, and 
to test other approximate functionals in the $\alpha\to\infty$ 
limit \cite{testAPPonSCE}.

Future work will be mainly devoted to the
 calculation of this limit for non-spherical densities, and to
study the generalization to the next leading term, $W'_\infty[\rho]$.
The calculation of the intracule density (the probability distribution
for the electron-electron distance) will be also carried out in connection
with the correlation energy functional constructed in Ref.~\onlinecite{Over}.

\section*{Acknowledgments}
We thank Prof. C. Bunge for the fully 
correlated density
of the Li atom, Prof. C.J Umrigar for the 
VMC densities of the Ne and Be atoms, and Dr. J. Toulouse for
the sphericalized densities of the B and C atoms. One of the authors (P.G.-G.)
gratefully acknowledges Prof. E.J. Baerends for the warm hospitality
in his group, where part of this work was done.

\appendix
\section{Harmonic interactions}
It is instructive to see how the SCE solution changes if we
replace the Coulomb repulsion with an harmonic interaction.
Since in this case we have the exact solution, we can also clarify 
how the SCE construction presented in this paper 
gives the $\alpha\to\infty$ limit of DFT.
We thus consider the exactly solvable  2-electron hamiltonian
\cite{davidson}
\beq
\hat{H}^\alpha=-\frac{1}{2}\left(\nabla^2_1+\nabla^2_2\right)
+\frac{k_\alpha}{2}  \left(r_1^2+r_2^2\right)-\frac{\alpha}{2}|\rv_1-\rv_2|^2,
\eeq
and we analyze the case in which $\alpha\to\infty$ and the density
is kept fixed, equal to the one at $\alpha=1$, by means of a suitable
$k_\alpha$.

The ground-state one-electron density of the hamiltonian $\hat{H}^\alpha$ 
is given by
\beq
\rho(r)=\frac{2 \beta^{3/2}}{\pi^{3/2}}\,e^{-\beta r^2},  \qquad 
\beta=\frac{2\sqrt{k_\alpha(k_\alpha-2\alpha)}}{\sqrt{k_\alpha-2\alpha}+
\sqrt{k_\alpha}},
\eeq
so that $k_\alpha$ must keep $\beta$ independent of $\alpha$ and thus satisfies the equation
\beq
\frac{\sqrt{k_\alpha(k_\alpha-2\alpha)}}{\sqrt{k_\alpha-2\alpha}+
\sqrt{k_\alpha}}=\frac{\sqrt{k_1(k_1-2)}}{\sqrt{k_1-2}+
\sqrt{k_1}},
\eeq
where $k_1>2$ in order to have a bound system. We easily find
\beq
k_{\alpha\to\infty}=2\alpha + \frac{\beta^2}{4} +O(\alpha^{-1/2}).
\label{eq_kinf}
\eeq
The square of the spatial wavefunction is equal to
\begin{eqnarray}
|\psi_\alpha(\rv_1,\rv_2)|^2  =  \left[\frac{\sqrt{k_\alpha(k_\alpha-2\alpha)}}{\pi^2}
\right]^{3/2} \times \nonumber \\
 e^{-\frac{1}{2}(\sqrt{k_\alpha}+\sqrt{k_\alpha-2\alpha})
(r_1^2+r_2^2)- (\sqrt{k_\alpha}-\sqrt{k_\alpha-2\alpha})\rv_1
\cdot\rv_2}.
\label{eq_psiexact}
\end{eqnarray}

Now, we want to compare the exact result with the SCE construction. We thus follow 
the same steps done for the case of the He atom in Subsec.~\ref{sub_He}. Again, we find
that the minimizing angle is $\theta=\pi$ and that the radial co-motion function must
satisfy the property $f(f(r))=r$, so that the only possible choices are the ones
of Eqs.~\eqref{breathing}-\eqref{increasing}. In this case, however,
the ``breathing'' solution of  Eq.~\eqref{breathing}, $f(r)=r$, is the one that yields 
the minimum
expectation of ``$\hat{V}\ee$'', and that is the 
global minimum of the 
corresponding $E\pot$. The very different nature of harmonic forces with respect to
Coulomb repulsion makes the solution completely different also from a qualitative point of
view. The ``breathing'' solution, that was a maximum for Coulomb forces, becomes the true minimum in the
case of harmonic forces. In this latter case, in fact, the interaction is not singular at zero
electron-electron distance so that the two electrons can get on top of each other (and they do,
at the ``nucleus''). This means that in this case even in the $\alpha\to\infty$ limit 
the spin state can still
play a role. The SCE one-body potential is obtained by
\beq
v'(r)=r+f(r)=2 r,
\eeq
so that $v(r)=r^2$ agrees with the leading term ($O(\alpha)$) of Eq.~\eqref{eq_kinf} (remember that
in the SCE construction we directly consider all the quantities divided by $\alpha$).

In order to compare the SCE solution with the exact wavefunction of Eq.~\eqref{eq_psiexact} 
in the $\alpha\to\infty$ limit, we have first to keep in mind that Eq.~\eqref{eq_psiexact} is
the solution of the hamiltonian $\hat{H}^\alpha$, so that it includes also the kinetic energy
operator $\hat{T}$. Equation \eqref{eq_kinf} shows that the external potential 
$\hat{V}\ext^\alpha$
that keeps the density fixed has the large-$\alpha$ expansion  $\hat{V}\ext^\alpha=\alpha \hat{V}_\infty
+\hat{V_0}+O(\alpha^{-1/2})$. The SCE construction only gives $\hat{V}_{\infty}$, but the 
exact solution of Eq.~\eqref{eq_psiexact} always includes also the next term, $\hat{V}_0$, because it is
of the same order of $\hat{T}$.  
The comparison with the SCE solution can be done in the following way.
Replace in Eq.~\eqref{eq_psiexact} the large-$\alpha$ expansion of $k_\alpha$ 
up to orders $\alpha^0$, $k_{\alpha}\to 2\alpha + \frac{\beta^2}{4}$. We can rearrange the
result, with $c=\frac{\beta^2}{4}$,
\begin{eqnarray}
|\psi_\alpha(\rv_1,\rv_2)|^2   \to   \left(\frac{\sqrt{2\alpha+c}-\sqrt{c}}{2 \pi}
\right)^{3/2} e^{-\frac{1}{2}(\sqrt{2\alpha+c}-\sqrt{c})|\rv_1+\rv_2|^2} \nonumber \\
 \times \frac{1}{\pi^{3/2}}\left(\frac{2\sqrt{c(2\alpha+c)}}{\sqrt{2\alpha+c}-\sqrt{c}}
\right)^{3/2} e^{-\sqrt{c}(r_1^2+r_2^2)}.
\label{eq_limit}
\end{eqnarray}
When $\alpha\to\infty$, the first gaussian in Eq.~\eqref{eq_limit} with its factor in front 
tends to the distribution $\delta(\rv_1+\rv_2)$. The second gaussian gives $\tfrac{1}{2}\rho(\rv_1)$. 
Thus, as $\alpha\to\infty$ we have
\beq
\lim_{\alpha\to\infty}|\psi_\alpha(\rv_1,\rv_2)|^2= \frac{\rho(\rv_1)}{2}\delta(\rv_1+\rv_2),
\eeq
which is exactly the SCE solution, since in the case of harmonic forces $\fv(\rv)=-\rv$.


\end{document}